\documentclass[twoside,slac_one]{revtex4}
\usepackage{graphicx}
\usepackage{fancyhdr}
\usepackage{amsmath} 
\usepackage{bm}
\usepackage{amsxtra}
\usepackage{amssymb}
\usepackage{amsthm}
\usepackage{latexsym}
\usepackage{lscape}

\newcommand{\met}{E_{\textrm{T}}^{\textrm{miss}}}

\begin{document}

\title{Search for High-Mass States with Lepton Plus Missing Transverse Energy Using the ATLAS Detector at Center-of-Mass Energy of 7 TeV}

%

\author{J. Degenhardt on behalf of ATLAS}
\affiliation{Department of Physics and Astronomy, University of Pennsylvania, Philadelphia, PA, USA}

\begin{abstract}
The ATLAS detector has been used to search for high-mass states decaying into a single high momentum lepton and missing transverse 
energy, such as new heavy charged gauge bosons. The latest search results for a $W^{\prime}$ boson decaying to lepton plus neutrino 
in $1.04$~fb$^{-1}$ of
proton-proton collisions at a center of mass energy of $7$~TeV produced at the Large Hadron Collider are presented.
\end{abstract}

\maketitle

\thispagestyle{fancy}


\section{Introduction}
In March of 2011 the Large Hadron Collider (LHC) had begun, once again, colliding protons at a center-of-mass energy of $7$~TeV.  By June the ATLAS detector had recorded more than $1$~fb$^{-1}$ of 
proton-proton collisions\footnote{The uncertainty assigned to the luminosity measurement for all channels is $3.7\%$.}.  These proceedings summarize the search for a $W^{\prime}$ boson decaying to a lepton (where lepton is defined as a muon or electron) plus a neutrino using $1.04$~fb$^{-1}$ of data 
collected by the ATLAS detector~\citep{atlas:wp-2011}.

There are many extensions to the Standard Model (SM) of particle physics that predict a new high-mass state similar to the $W$ boson in the SM~\citep{pdg:2010,ssm}.  The sequential, standard model (SSM) boson, i.e. the extended gauge boson of Ref.~\citep{ssm}, is used to model a $W^{\prime}$ boson with the same couplings to leptons as the SM $W$ boson but at a higher mass.

No $W^{\prime}$ boson signal is observed, and the data is used to set new limits~\citep{atlas:wprime_2010_pub} on $\sigma B$ (cross section times branching fraction) as a function of the $W^{\prime}$ boson mass.   Table~\ref{tab:xsec} shows the signal production cross sections used for each mass point.  The search is performed in the individual electron and muon channels where limits on $W^{\prime}\rightarrow e\nu$ and $W^{\prime}\rightarrow \mu\nu$ are set.  The channels are also combined assuming the same branching fraction for both channels.  The limits are determined by examining the transverse mass ($m_{\textrm{T}}$) distribution for 
resonant states above $500$~GeV, where $m_{\textrm{T}}$ is constructed from events with a single high transverse momentum ($p_{\textrm{T}}$) lepton plus large missing transverse energy ($\met$),
\begin{equation}
m_{\textrm{T}} = \sqrt{2p_{\textrm{T}} ~\met(1 - \cos \phi_{\ell\nu})}.
\label{eqn:mt}
\end{equation}
Here $\phi_{\ell\nu}$ is the angle between the lepton $p_{\textrm{T}}$ and $\met$ vectors \footnote{The nominal interaction point is defined as the origin of the coordinate system, while the anti-
clockwise beam direction defines the $z$-axis and the $x$ -- $y$ plane is transverse to the beam direction.  The positive $x$-axis is defined as pointing from the interaction point to the center 
of the LHC ring and the positive $y$-axis is defined as pointing upwards.  The azimuthal angle $\phi$ is measured around the beam axis and the polar angle $\theta$ is the angle from the 
beam axis.  The pseudorapidity is defined as $\eta = -\ln \tan(\theta / 2)$.  The distance $\Delta R$ in $\eta$ -- $\phi$ space is defined as 
$\Delta R = \sqrt{(\Delta\eta)^{2} + (\Delta\phi)^{2}}$.}.  

 A bayesian method is used to  search for an excess of the $m_{\textrm{T}}$ signal above the expected backgrounds.  When no excess is observed (above $3\sigma$) a limit on the 
cross section ($\sigma_{\textrm{SSM}}$ significance) times branching ratio ($BR$) is set where the SSM $W^{\prime}$ boson is used as a benchmark for setting limits.

These proceedings begin with a brief description of the ATLAS detector in section \ref{sec:detector}, followed by an explanation of the event selection in section \ref{sec:evtsel}.  The explanation of the  
modeling of the backgrounds is done in section \ref{sec:bknds}.  Section \ref{sec:results} presents the final results and these proceedings are concluded in Section \ref{sec:conclusion}.
\section{Description of the ATLAS Detector \label{sec:detector}}
The ATLAS detector is a general purpose particle detector that is characterized by three different components, the inner-detector (ID), the calorimeter, and the muon system (MS).  The innermost 
system is the inner-detector which consists of a silicon pixel detector and a silicon strip detector covering $|\eta| < 2.5$.  A transition radiation tracker is the outer tracker of the ID and covers $|\eta| < 2.0$.  The ID is immersed in a homogenous $2$~T field produced from a 
superconducting solenoid.  The ID allows reconstruction of the vertices and tracks of charged particles originating from the interaction region.  A finely segmented, hermetic calorimeter system surrounds the ID and covers $|\eta| < 4.9$ as well as provides three-dimensional reconstruction of particle showers.   The electromagnetic (EM) calorimeter uses liquid argon, while the outer calorimeter, which primarily detects hadronic showers,  uses scintillating tiles in the barrel ($|\eta| < 1.7$) and liquid argon for higher $|\eta|$.  The energy resolution for an electron approaches $1\%$ for electrons above $50$~GeV in $E_{\textrm{T}}$.  The MS, situated outside the calorimeters, is 
immersed in a $\approx 1.4$~T toroidal field that is produced by superconducting air-core toroids.  The integral magnetic field for the MS is about $3$~Tm on average.  The curvature of the muons as they pass through the magnetic field is measured with three layers of precision drift-tube chambers in the region $|\eta| < 2.0$ and one layer of cathode-strip chambers followed by two layers of drift-tube chambers for $2.0 < |\eta| < 2.7$.  Resistive-plate and thin-gap chambers provide the triggering for muons and measure the $\phi$ coordinate.  The muon system is able to achieve a momentum resolution of $\approx 15\%$ at $1$~TeV.

\section{Event Selection \label{sec:evtsel}}
The event selection for this analysis first requires that all events were recorded while the detector was operating in an optimal fashion.  A well reconstructed primary vertex, with at 
least three reconstructed tracks associated with the vertex is required to be present in the event (all objects selected in the event are then required to be associated with this primary vertex).  The primary 
vertex is also required to be reconstructed within $200$~mm, along the $z$-axis, of the center of the interaction region.  Events are then examined to determine if any jets have abnormal shapes.  
If an event has a poorly reconstructed jet, then it is thrown out as this type of event could bias the $\met$ measurement.  Only events that fire specific single-lepton triggers are 
included in the search.  For the electron channel the lowest unprescaled lepton trigger for the dataset has a $22$~GeV $E_{\textrm{T}}$ threshold.  The trigger requires an EM cluster in 
the EM calorimeter  that has an energy deposition profile consistent with that of an electron.  The trigger also requires that the EM cluster be matched to a track in the inner 
detector.  The muon trigger has to be fired by the RPC or the TGC systems, and to pass a $p_{\textrm{T}}$ threshold of $20$~GeV.  Then the trigger object has to match an inner 
detector track.

In the electron channel, the reconstructed electron energy must be larger than $25$~GeV and the electron must have an $|\eta| < 1.37$ or $1.52 < |\eta| < 2.47$.  
The offline energy cluster must have an energy deposit that is consistent with that of an electron and is matched to an offline track.  The offline track is required to have a transverse impact parameter 
within $0.1$~cm of the primary vertex and a longitudinal distance from the primary vertex within $1.0$~cm.  A hit from the first layer of the pixel detector is required to be associated with the 
electron track.  Finally the energy isolation of the electron, $\Sigma E_{\textrm{T} \textrm{Calo}} (R < 0.4)$ is required to be less than $9$~GeV.  Events with an additional electron are vetoed.

The muon channel makes the following offline requirements.  The offline muon track is first matched to the trigger object.  A $p_{\textrm{T}}$ cut of $25$~GeV is then required, where the $p_{\textrm{T}}$ is 
the combined transverse momentum in the MS and the ID.  The $|\eta|$ of the muon must be less than $1.0$ or $1.3 < |\eta| < 2.0$.  The combined track of the muon is required to have a transverse 
impact parameter less than $0.1$~cm from the primary vertex and a longitudinal distance from the primary vertex less than $1.0$~cm.  The muon track in the MS is required to register hits in all 
three stations of the MS, ensuring the best possible momentum measurement.  As an additional quality cut, the significance of the difference of the curvature measurements between the ID and the MS is required to be less than $5\sigma$.  This requirement ensures that no poorly measured muons pass into the final sample.  The track 
isolation, $\Sigma p_{\textrm{T}} (\textrm{ID tracks } R<0.3)$, is to be less than $5\%$ of the muon's $p_{\textrm{T}}$.  Events with additional muons are also vetoed.

\begin{figure}[ht]
\includegraphics[width=80mm]{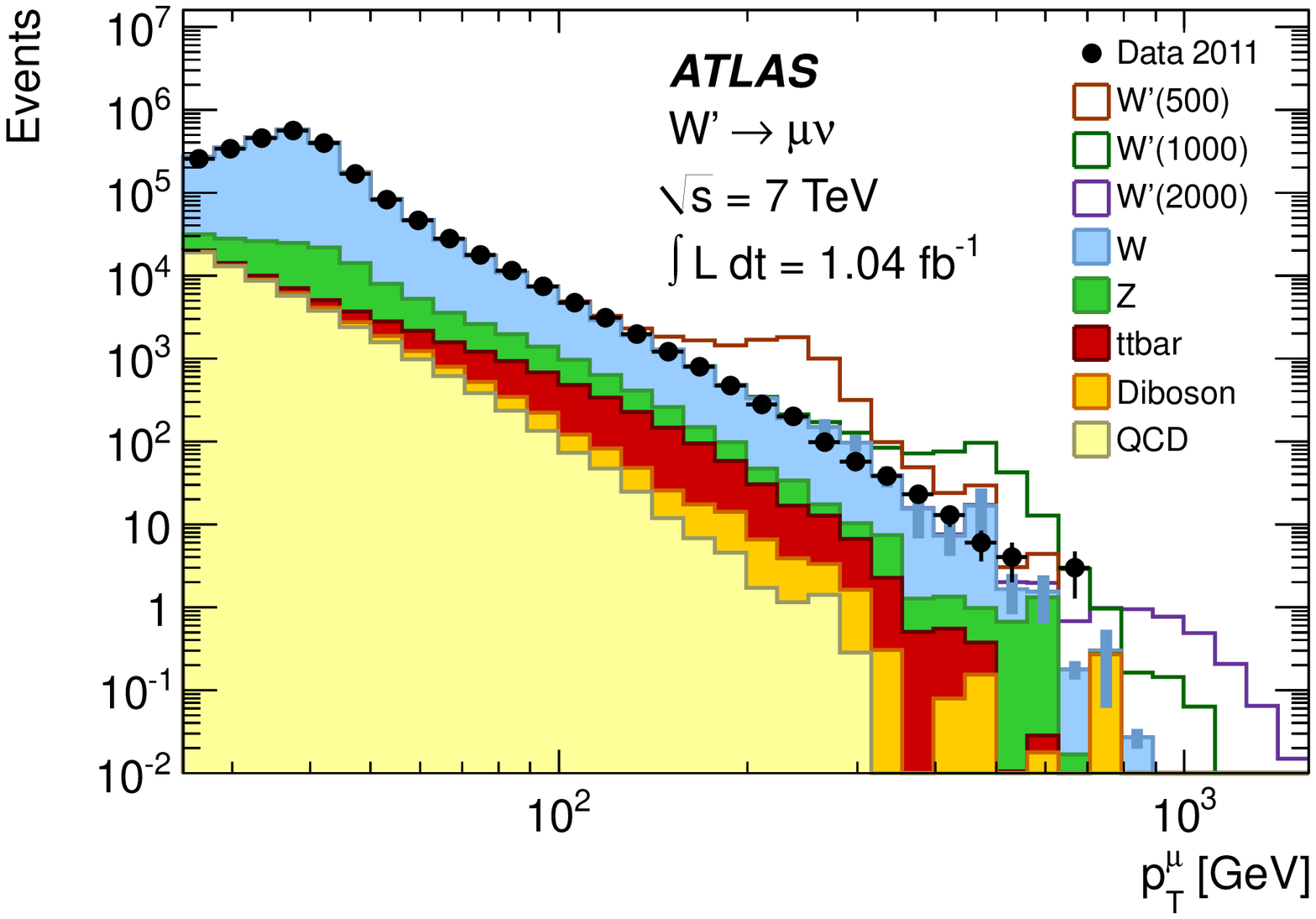}%
\includegraphics[width=80mm]{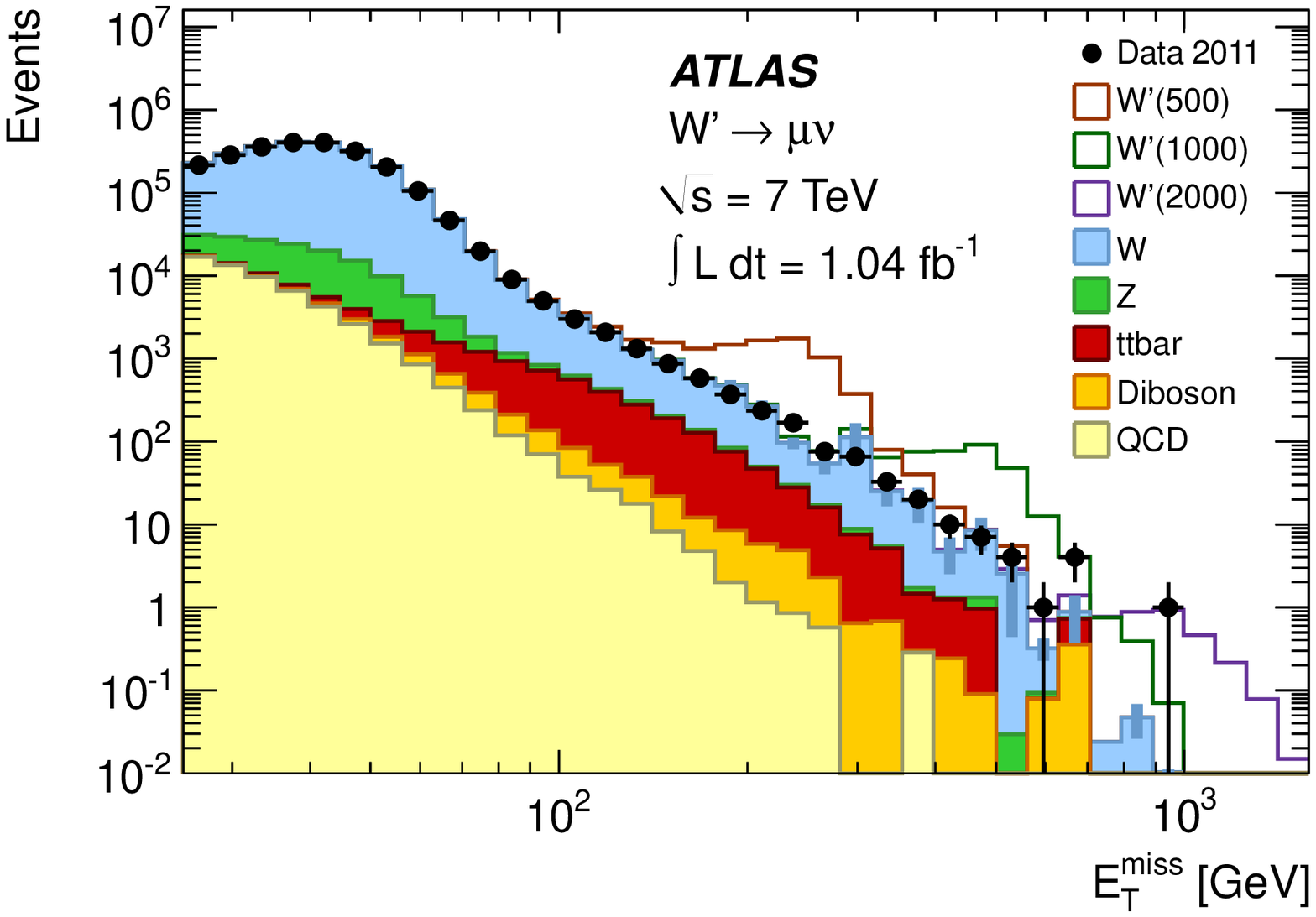}%
\caption{Lepton $p_{\textrm{T}}$(left) and $\met$(right) spectra of the muon channel after event selection. The points represent data and the filled histograms show the stacked backgrounds.  Open histograms are $W^{\prime}\rightarrow \mu \nu$ signals added to the background with masses in GeV indicated in parentheses in the legend.  The QCD backgrounds estimated from data are also shown.  The signal and other background samples are normalized using the integrated luminosity of the data and the NNLO (approximate-NNLO for $t\bar{t}$) cross sections listed in Table~\ref{tab:xsec} \label{fig:mu-distributions}.}
\end{figure}

\begin{figure}[ht]
\includegraphics[width=80mm]{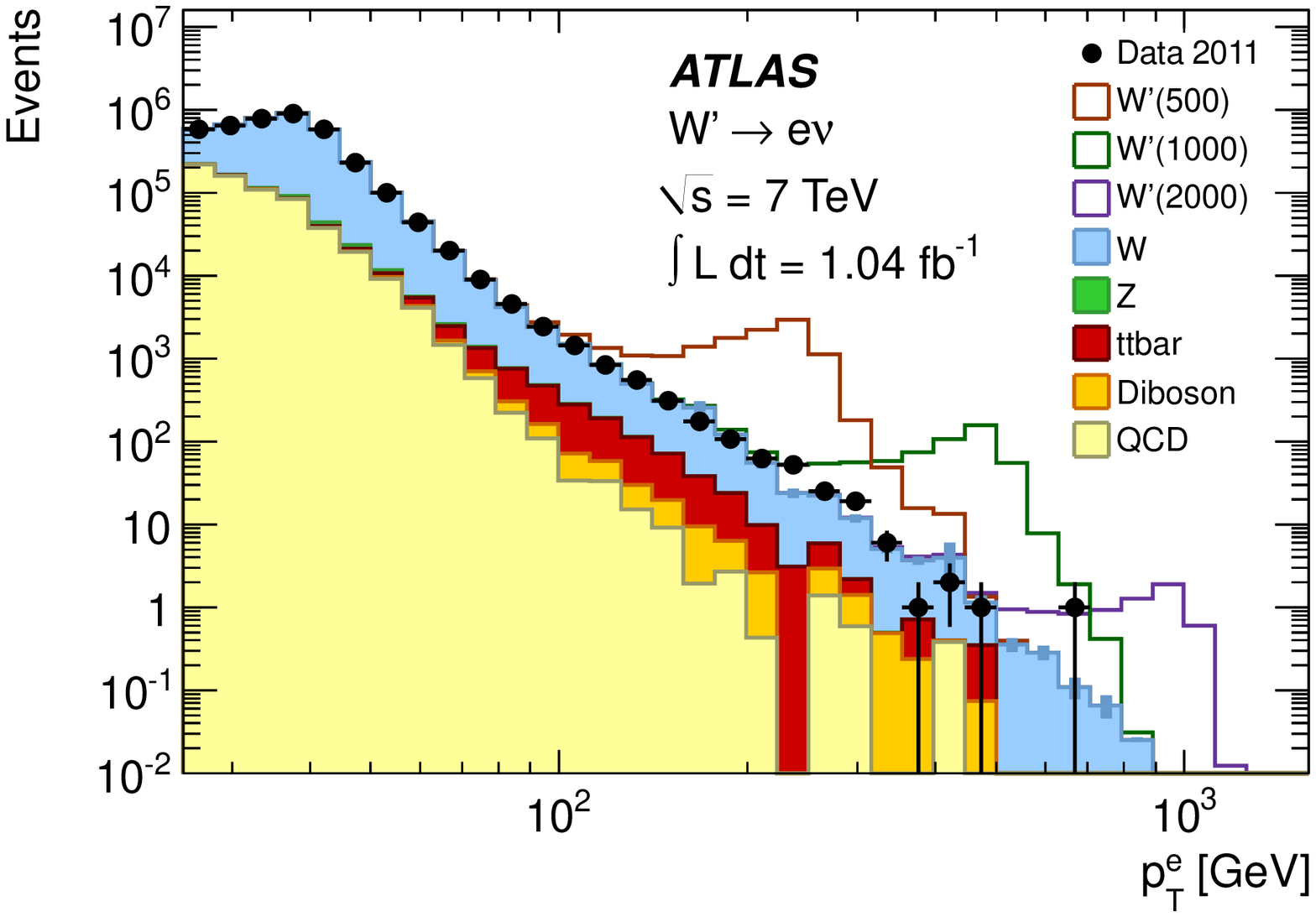}%
\includegraphics[width=80mm]{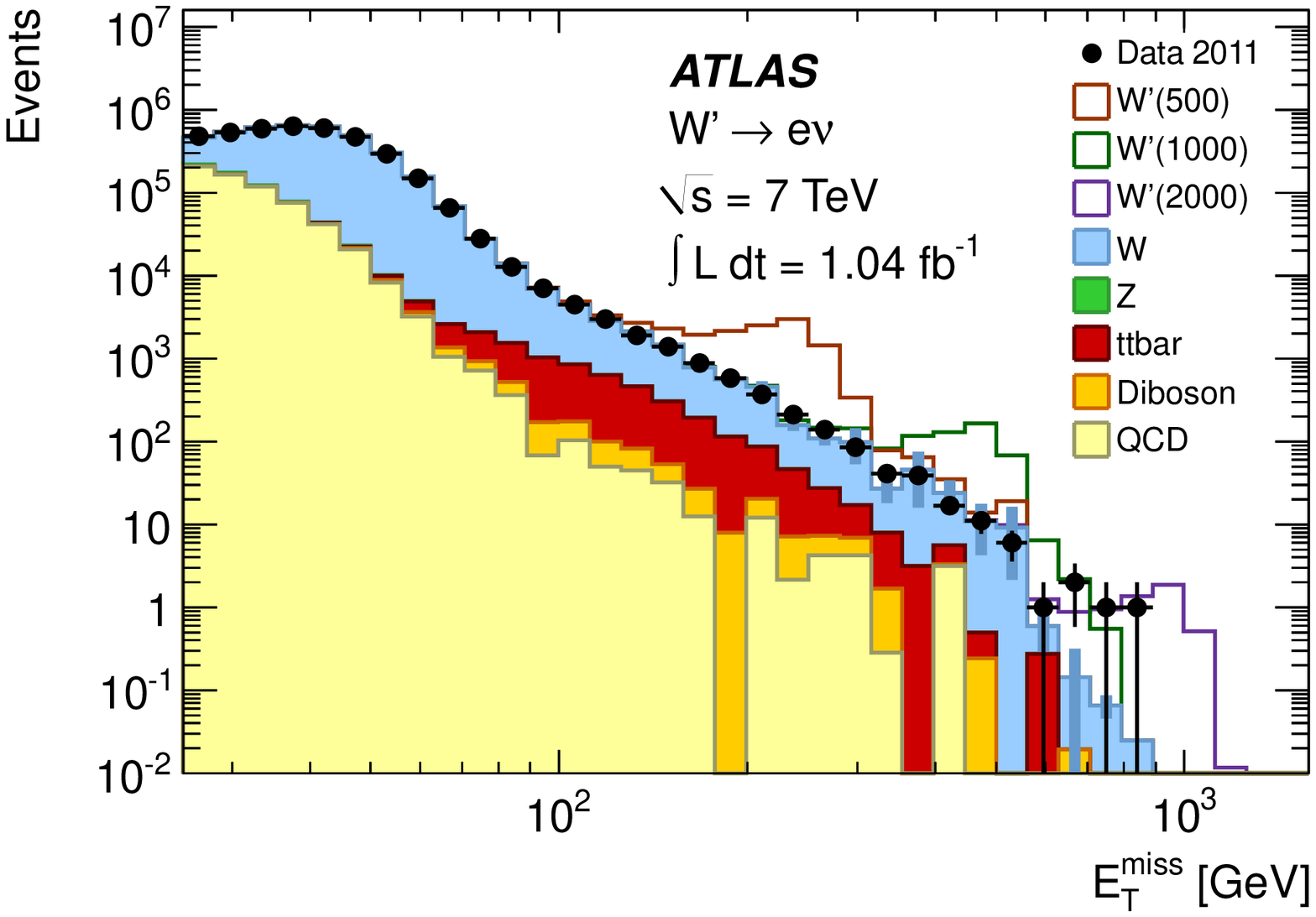}%
\caption{Lepton $p_{\textrm{T}}$ (left) and $\met$(right) spectra of the electron channel after event selection. The points represent data and the filled histograms show the stacked backgrounds.  Open histograms are $W^{\prime}\rightarrow e \nu$ signals added to the background with masses in GeV indicated in parentheses in the legend.  The QCD backgrounds estimated from data are also shown.  The signal and other background samples are normalized using the integrated luminosity of the data and the NNLO (approximate-NNLO for $t\bar{t}$) cross sections listed in Table~\ref{tab:xsec}.\label{fig:el-distributions}}
\end{figure}

Finally both channels are required to have $\met$ greater than $25$~GeV.  In order to further reduce the amount of background from light flavor or heavy flavor decays in the electron channel, the 
$\met$ is required to be larger than $60\%$ of the electron $E_{\textrm{T}}$.    The final distributions of the lepton $p_{\textrm{T}}$ and 
$\met$ are shown in Figs.~\ref{fig:mu-distributions} and \ref{fig:el-distributions}. These figures also show several mass point distributions of SSM $W^{\prime}$ boson MC signal.  The SSM $W^{\prime}$ boson production MC has been generated using {\sc pythia} and corrected for NNLO QCD corrections.

\begin{table}[tb]
\caption{
Calculated values of $\sigma B$ for $W^{\prime}\rightarrow \ell \nu$ and the leading backgrounds.
The value for $t\bar{t}\rightarrow\ell X$ includes all final states with at least one
lepton ($e$, $\mu$ or $\tau$).
The other processes are exclusive and are used for both $\ell=e$ and $\ell=\mu$.
All calculations are NNLO except $t\bar{t}$ which is approximate-NNLO.
}
\label{tab:xsec}
\begin{center}
\begin{tabular}{l|c|l}
\hline
\hline
        & Mass        &             \\
Process & [GeV] &  $\sigma B$ [pb] \\
\hline
$W^{\prime}\rightarrow \ell \nu$ &
   \phantom{0}500 & \phantom{000}17.25    \\
 & \phantom{0}600 & \phantom{0000}8.27    \\
 & \phantom{0}750 & \phantom{0000}3.20    \\
 & \phantom{}1000 & \phantom{0000}0.837   \\
 & \phantom{}1250 & \phantom{0000}0.261   \\
 & \phantom{}1500 & \phantom{0000}0.0887  \\
 & \phantom{}1750 & \phantom{0000}0.0325  \\
 & \phantom{}2000 & \phantom{0000}0.0126  \\
 & \phantom{}2250 & \phantom{0000}0.00526 \\
 & \phantom{}2500 & \phantom{0000}0.00234 \\
\hline
$W\rightarrow \ell \nu$ & & \phantom{}10460 \\
$Z/\gamma^{*}\rightarrow \ell \ell$               & & {\phantom{00}989} \\
($m_{Z/\gamma^{*}}>60$~GeV)  & & \\
 $t\bar{t} \rightarrow \ell X$ & &  \phantom{000}89.4 \\
\hline
\hline
\end{tabular}
\end{center}
\end{table}

\begin{figure}[ht]
\includegraphics[width=80mm]{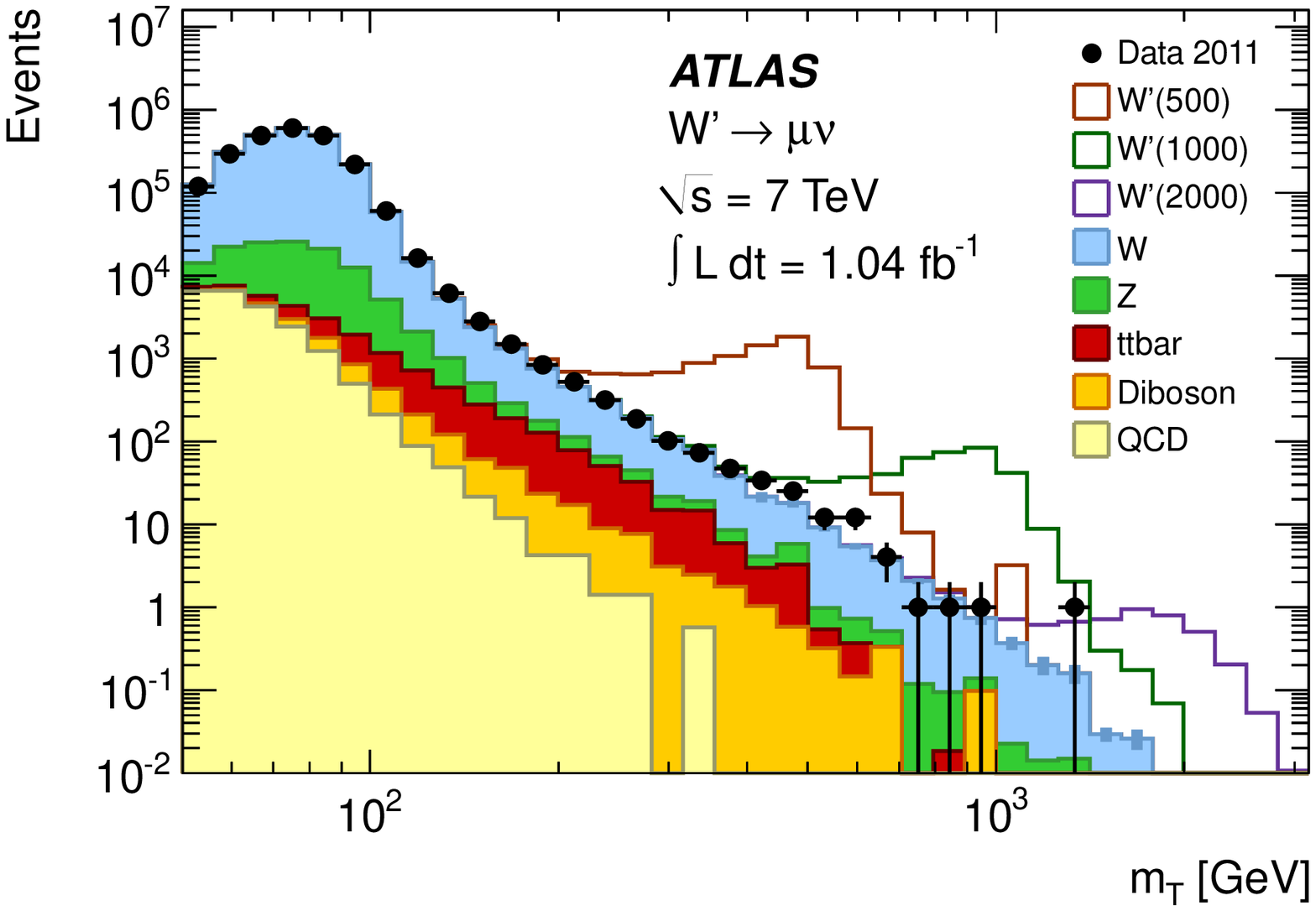}%
\includegraphics[width=80mm]{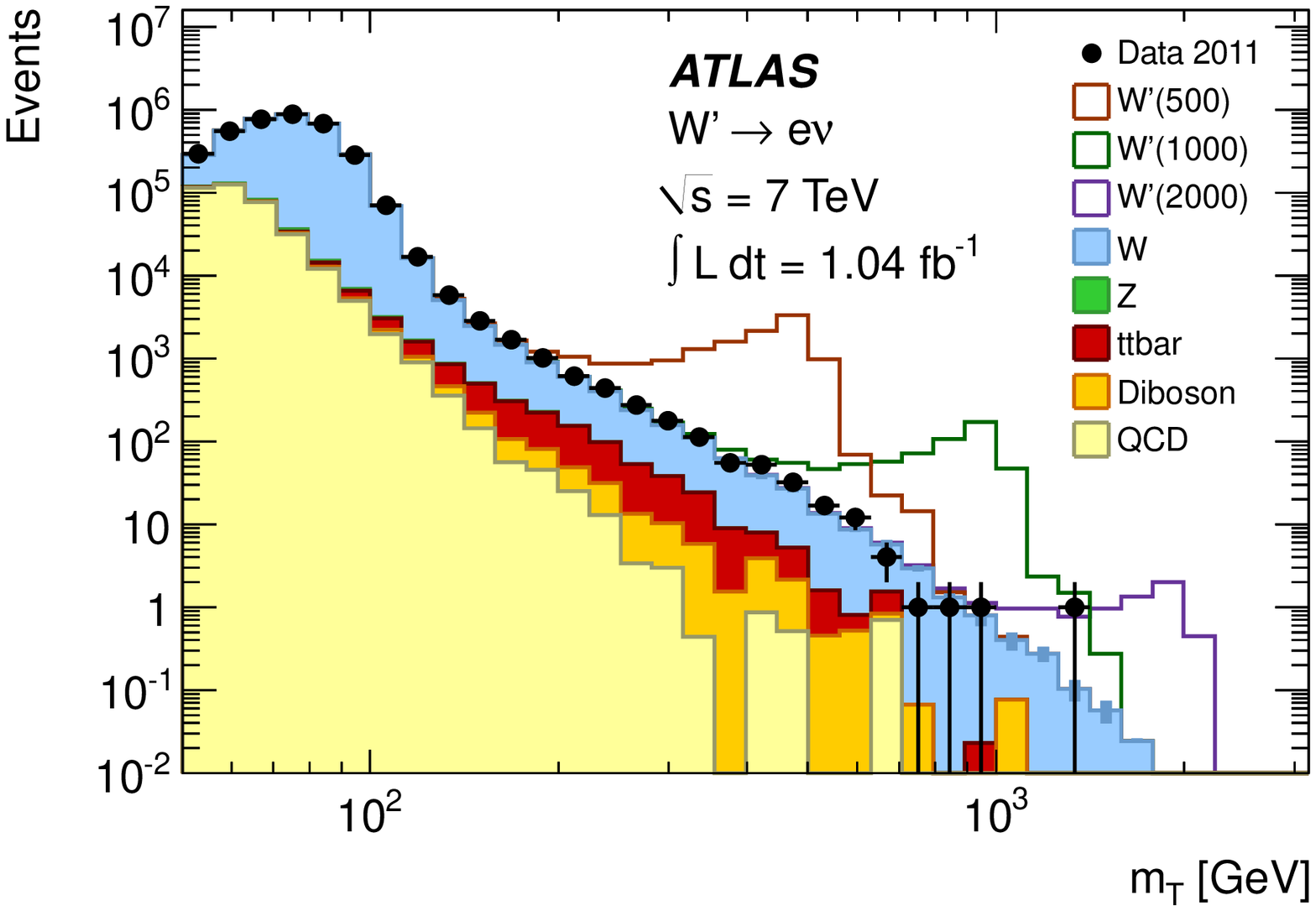}%
\caption{Event $m_{\textrm{T}}$ spectra of the muon (left) and electron (right) channels after event selection. The points represent data and the filled histograms show the stacked backgrounds.  Open histograms are $W^{\prime}\rightarrow \ell \nu$ signals added to the background with masses in GeV indicated in parentheses in the legend.  The QCD backgrounds estimated from data are also shown.  The signal and other background samples are normalized using the integrated luminosity of the data and the NNLO (approximate-NNLO for $t\bar{t}$) cross sections listed in Table~\ref{tab:xsec}. \label{fig:mt-distributions}}
\end{figure}

\section{Background Modeling \label{sec:bknds}}
The largest contribution to the background is the SM $W$ and $Z$ boson production.  The SM $W$ boson is an irreducible 
background and is modeled using {\sc pythia} to generate the background events at LO with the shape of the $m_{\textrm{T}}$ distribution then corrected to NNLO.  The $Z$ boson production becomes a contribution when one of the leptons from the leptonic decay is not reconstructed and then mimics real $\met$.  This contribution is reduced by requiring an additional lepton veto in the event, and is modeled in the simulation.  

The next largest contributions to the backgrounds are SM diboson and $t\bar{t}$ production.  These events also have real leptons and 
$\met$ and are modeled using simulated events.  The $t\bar{t}$ production is corrected to approximate-NNLO~\citep{Moch:2008qy, Langenfeld:2009tc, Aliev:2010zk}.  The production cross sections used are given in Tab.~\ref{tab:xsec}.

The last significant contribution to the backgrounds are events where a hadron, or photon, will mimic an electron or 
produce a muon in the course of the decay of the hadron.  These types of events, when produced with sufficient $\met$ will pass the selection cuts of the analysis. Such background events are collectively referred to as 
QCD production.  This contribution ends up being negligible as it is very rare for an isolated muon to be produced 
from a hadronic decay.  To model this contribution to the electron channel, an ABCD method was used~\citep{atlas:photon2010}, where the discriminating variables used were $\met$ and the electron isolation.  The fraction of isolated electrons 
was determined in the low $\met$ region and extrapolated to the high $\met$ region.  Table~\ref{tab:nbg} shows the contributions of the backgrounds for a particular mass point.

\begin{table}[!bt]
\caption{
Expected numbers of events from the various background
sources in each decay channel
for $m_{\textrm{T}}>891$~GeV, the region used to
search for a $W^{\prime}$ boson with a mass of $1500$~GeV.
The $W\rightarrow \ell \nu$ and $Z \rightarrow \ell \ell$ entries include the expected contributions from the $\tau$-lepton.
No muon events are found in the $t\bar{t}$ sample above this $M_{\textrm{T}}$ threshold.
The uncertainties are statistical.  The final two rows shows the number of expected signal events ($N_{\textrm{sig}}$) and the number of observed events ($N_{\textrm{obs}}$) in each channel.  The uncertainties on $N_{\textrm{sig}}$ include systematic uncertainties except that of the luminosity.
}
\label{tab:nbg}
\begin{center}
\begin{tabular}{l | l | l}
\hline
\hline
                & \multicolumn{1}{c|}{$e\nu$} & \multicolumn{1}{c}{$\mu\nu$} \\
\hline
$W\rightarrow \ell \nu$           & \phantom{0}1.59\phantom{000} $\pm$ 0.13         & \phantom{0}1.36\phantom{0}  $\pm$ 0.13 \\
$Z\rightarrow \ell \ell$            &\phantom{0} 0.00010\phantom{} $\pm$ 0.00004      & \phantom{0}0.095\phantom{}  $\pm$ 0.005 \\
diboson         & \phantom{0}0.08\phantom{000} $\pm$ 0.08         & \phantom{0}0.11\phantom{0}  $\pm$ 0.08 \\
$t\bar{t}$          & \phantom{0}0.08\phantom{000} $\pm$ 0.08         & \phantom{0}0 \\
QCD             & \phantom{0}0\phantom{.00000} $^{+0.17}_{-0}$    & \phantom{0}0.01\phantom{0} $^{+0.02}_{-0.01}$ \\
\hline
Total           & \phantom{0}1.75\phantom{000} $^{+0.24}_{-0.18}$ & \phantom{0}1.57\phantom{0} $\pm$ 0.15 \\
\hline
$N_{\textrm{sig}}$ & \phantom{0} 49.6 $\pm$ 6.0 \phantom{0} & \phantom {0} 34.4 $\pm$ 4.4\phantom{0} \\
\hline
$N_{\textrm{obs}}$ &  \phantom{000000} $2$ \phantom{0} & \phantom{00000} $2$ \phantom{0} \\
\hline
\hline
\end{tabular}
\end{center}
\end{table}

\section{Results \label{sec:results}}
To determine the significance of excess data of any events above the minimum $m_{\textrm{T}}$ ($m_{\textrm{T}~\textrm{min}}$), a bayesian method is used.  The $m_{\textrm{T min}}$ is determined by minimizing the expected limit on $\sigma B$ in a given mass bin.   
First a likelihood is constructed from the expected number of events, $N_{\textrm{exp}}$, where,  
\begin{equation}
N_{\textrm{exp}} = \epsilon_{\textrm{sig}} L_{\textrm{int}} \sigma B + N_{\textrm{bg}}.
\end{equation}
Here, $\epsilon_{\textrm{sig}}$ is the signal efficiency, $L_{\textrm{int}}$ is the integrated luminosity, $\sigma B$ is the signal production cross section times branching ratio, and $N_{\textrm{bg}}$ is the number of background events.  The Poisson statistics likelihood, $\cal L$ to observe $N_{\textrm{obs}}$ events is 
\begin{equation}
  {\cal L}(N_{\textrm{obs}}|\sigma B) = \frac{(\epsilon_{\textrm{sig}} L_{\textrm{int}} \sigma B + N_{\textrm{bg}})^{N_{\textrm{obs}}} e^{-(\epsilon_{\textrm{sig}} L_{\textrm{int}} \sigma B + N_{\textrm{bg}})}}{N_{\textrm{obs}}!}.
\label{eqn:lhood}
\end{equation}
Systematic uncertainties are treated as Gaussian nuisance parameters, $\theta_{i}$, in the likelihood, 
\begin{equation}
  {\cal L}(N_{\textrm{obs}}|\sigma B, \theta_{i}) =  \int  {\cal L}(N_{\textrm{obs}}|\sigma B)   \prod g_{i}(\theta_{i})d\theta_{i},
\label{eqn:lhoodn}
\end{equation}
where $g_{i}(\theta_{i})$ is a probability density function (pdf) for a given nuisance parameter.  The main contributions to the nuisance parameters arise from $L_{\textrm{int}},~\epsilon_{\textrm{sig}}$, and $N_{\textrm{bg}}$.  These values are given in Table~\ref{tab:syst_summary}.  The correlations between the nuisance parameters are neglected given their small effect on the limits. 

\begin{table*}[ht]
\caption{
  Relative uncertainties on the event selection efficiency and background level for
  a $W^{\prime}$ boson with a mass of 1500~GeV.
  The efficiency uncertainties include contributions from trigger, reconstruction and event selection.
  The cross section uncertainty for $\epsilon_{\textrm{sig}}$ is that assigned to the acceptance correction due to the differences in the acceptance from the LO generator and the NNLO.  These uncertainties vary from $7\%$ at a $500$~GeV $W^{\prime}$ mass to $10\%$ at a $2500$~GeV mass.
  The last row gives the total uncertainties.
\label{tab:syst_summary}
}
\begin{center}
\begin{tabular}{l|rr|rr}
\hline
\hline
 & \multicolumn{2}{c|}{$\epsilon_{\textrm{sig}}$} & \multicolumn{2}{c}{$N_{\textrm{bg}}$} \\
 Source                      &  \multicolumn{1}{c}{$e\nu$}  & \multicolumn{1}{c|}{$\mu\nu$} &  \multicolumn{1}{c}{$e\nu$}  & \multicolumn{1}{c}{$\mu\nu$} \\
\hline
 Efficiency                  &       2.7\%  &       3.9\%  &       2.7\%  &       3.8\%  \\
 Energy/momentum resolution  &       0.3\%  &       2.3\%  &       2.9\%  &       0.6\%  \\
 Energy/momentum scale       &       0.5\%  &       1.3\%  &       5.2\%  &       3.0\%  \\
 QCD background              & \multicolumn{1}{c}{-} & \multicolumn{1}{c|}{-} &
                                                                  10.0\%  &       1.3\%  \\
 Monte Carlo statistics      &       2.5\%  &       3.1\%  &       9.4\%  &       9.9\%  \\
 Cross section (shape/level) &       3.0\%  &       3.0\%  &       9.5\%  &       9.5\%  \\
\hline
 All                         &       4.7\%  &       6.3\%  &      18\%\phantom{.0}  &      15\%\phantom{.0}  \\
\hline
\hline
\end{tabular}
\end{center}
\end{table*}

Using this likelihood a posterior probability, $P_{\textrm{post}}(\sigma B)$, is constructed 
\begin{equation}
P_{\textrm{post}}(\sigma B) = N {\cal L}(\sigma B) P_{\textrm{prior}}(\sigma B),
\label{eqn:post}
\end{equation}
where $P_{\textrm{prior}}(\sigma B)$ is the canonical flat prior probability, and $N$ is a normalization factor so that 
the integrated probability is equal to one.  The posterior probability is evaluated for each mass and each decay 
channel as well as their combinations.  Then the posterior probabilities are used to assess the discovery significance and set a limit on $\sigma B$.
No excess above $3\sigma$ significance is observed, so $95\%$ confidence level (CL) limits on $\sigma B$ are set.  
The inputs to the likelihood are shown in Tab.~\ref{tab:limit_input}.  
\begin{table*}[ht]
\caption{Inputs for the $W^{\prime}\rightarrow e\nu$ and $W^{\prime}\rightarrow \mu\nu~\sigma B$ limit calculations.
The first three columns are the $W^{\prime}$ boson mass, $m_{\textrm{T}}$ threshold and decay channel.
The next two are the signal selection efficiency, $\epsilon_{\textrm{sig}}$, and the prediction
for the number of signal events, $N_{sig}$, obtained with this efficiency.
The last two columns are the expected number of background
events, $N_{\textrm{bg}}$, and the number of events observed in data, $N_{\textrm{obs}}$.
The uncertainties on $N_{\textrm{sig}}$ and $N_{\textrm{bg}}$ include contributions from the uncertainties on the
cross sections but not from the integrated luminosity.
\label{tab:limit_input}
}
\begin{center}
\begin{tabular}{rrr| r@{ $\pm$ }r r@{ $\pm$ }r r@{ $\pm$ }r r}
\hline
\hline
 $m_{W^{\prime}}$   & $m_{\textrm{T} \textrm{min}}$ & \\
~[GeV] & [GeV] &     & \multicolumn{2}{c}{$\epsilon_{\textrm{sig}}$} & \multicolumn{2}{c}{$N_{\textrm{sig}}$} & \multicolumn{2}{c}{$N_{\textrm{bg}}$} & $N_{\textrm{obs}}$ \\
\hline
\hline
 500 & 398 & $e\nu$   & 0.388 & 0.019 & 6930\phantom{.00} & 620\phantom{.00}  & 101.9\phantom{00} & 10.8\phantom{00}  & 121 \\
 500 & 398 & $\mu\nu$ & 0.252 & 0.015 & 4500\phantom{.00} & 430\phantom{.00}  &  63.7\phantom{00} &  6.5\phantom{00}  &  91 \\
\hline
 600 & 447 & $e\nu$   & 0.456 & 0.022 & 3910\phantom{.00} & 330\phantom{.00}  &  62.1\phantom{00} &  7.1\phantom{00}  &  69 \\
 600 & 447 & $\mu\nu$ & 0.286 & 0.016 & 2450\phantom{.00} & 220\phantom{.00}  &  41.8\phantom{00} &  4.7\phantom{00}  &  57 \\
\hline
 750 & 562 &   $e\nu$ & 0.429 & 0.020 & 1420\phantom{.00} & 110\phantom{.00}  &  20.7\phantom{00} &  3.7\phantom{00}  &  20 \\
 750 & 562 & $\mu\nu$ & 0.293 & 0.017 &  970\phantom{.00} &  79\phantom{.00}  &  14.3\phantom{00} &  1.4\phantom{00}  &  20 \\
\hline
1000 & 708 &   $e\nu$ & 0.482 & 0.022 &  417\phantom{.00} &  35\phantom{.00}  &   6.13\phantom{0} &  0.92\phantom{0}  &   4 \\
1000 & 708 & $\mu\nu$ & 0.326 & 0.019 &  282\phantom{.00} &  26\phantom{.00}  &   4.98\phantom{0} &  0.54\phantom{0}  &   4 \\
\hline
1250 & 794 &   $e\nu$ & 0.527 & 0.024 &  143\phantom{.00} &  14\phantom{.00}  &   3.09\phantom{0} &  0.49\phantom{0}  &   3 \\
1250 & 794 & $\mu\nu$ & 0.367 & 0.021 &   99\phantom{.00} &  10\phantom{.00}  &   2.87\phantom{0} &  0.34\phantom{0}  &   3 \\
\hline
 1500 & 891 &   $e\nu$ & 0.541 & 0.026 &   49.6\phantom{0} &   6.0\phantom{0}  &   1.75\phantom{0} &  0.32\phantom{0}  &   2 \\
 1500 & 891 & $\mu\nu$ & 0.374 & 0.024 &   34.4\phantom{0} &   4.4\phantom{0}  &   1.57\phantom{0} &  0.23\phantom{0}  &   2 \\
\hline
1750 & 1000 &   $e\nu$ & 0.515 & 0.024 &   17.3\phantom{0} &   2.4\phantom{0}  &   0.89\phantom{0} &  0.20\phantom{0}  &   1 \\
1750 & 1000 & $\mu\nu$ & 0.338 & 0.020 &   11.4\phantom{0} &   1.7\phantom{0}  &   0.82\phantom{0} &  0.14\phantom{0}  &   1 \\
\hline
2000 & 1122 &   $e\nu$ & 0.472 & 0.023 &    6.16\phantom{} &   0.99\phantom{}  &   0.48\phantom{0} &  0.10\phantom{0}  &   1 \\
2000 & 1122 & $\mu\nu$ & 0.323 & 0.021 &    4.21\phantom{} &   0.70\phantom{}  &   0.44\phantom{0} &  0.09\phantom{0}  &   1 \\
\hline
2250 & 1122 &   $e\nu$ & 0.415 & 0.019 &    2.84\phantom{} &   0.50\phantom{}  &   0.48\phantom{0} &  0.10\phantom{0}  &   1 \\
2250 & 1122 & $\mu\nu$ & 0.288 & 0.018 &    1.97\phantom{} &   0.36\phantom{}  &   0.44\phantom{0} &  0.09\phantom{0}  &   1 \\
\hline
2500 & 1122 &   $e\nu$ & 0.333 & 0.018 &    0.81\phantom{} &   0.16\phantom{}  &   0.48\phantom{0} &  0.10\phantom{0}  &   1 \\
2500 & 1122 & $\mu\nu$ & 0.221 & 0.017 &    0.53\phantom{} &   0.11\phantom{}  &   0.44\phantom{0} &  0.09\phantom{0}  &   1 \\
\hline
\hline
\end{tabular}
\end{center}
\end{table*}
Table~\ref{tab:limits_xbr} shows the upper limits on $W^{\prime}\rightarrow \ell \nu ~ \sigma B$ for the individual channels and different mass points.
\begin{table}[ht]
\caption{Upper limits on $W^{\prime}\rightarrow \ell \nu ~ \sigma B$.
         The first column is the mass bin and the following columns are
         the 95\% CL limits of the electron channel, the muon channel, and the combined limit respectively.  The limits include all of the systematics as described in the text.
\label{tab:limits_xbr}
}
\begin{center}
\begin{tabular}{c|ccc}
\hline
\hline
 $m_{W^{\prime}}$     & \multicolumn{3}{c}{95\% CL limit on $\sigma B$ [fb]} \\
 ~[GeV] &               $e\nu$ & $\mu\nu$ & both \\
\hline
500 & 121 & 191 & 130  \\
600 &  61  & 110  & 65 \\ 
750 &  28.5  & 51.7  & 28.1 \\ 
1000 & 10.6 & 16.7 & 7.7 \\ 
1250 &  10.1  & 14.7  & 7.5\\ 
1500 &  9.0  & 13.3  & 6.7 \\ 
1750 &   7.9 & 12.2  & 5.7 \\ 
2000 &   9.1 & 13.4  & 6.7 \\ 
2250 &   10.3 &  15.0 & 7.6 \\ 
2500 &  12.9  & 19.7  & 9.6 \\   

\hline
\hline
\end{tabular}
\end{center}
\end{table}

\begin{figure}
\includegraphics[width=80mm]{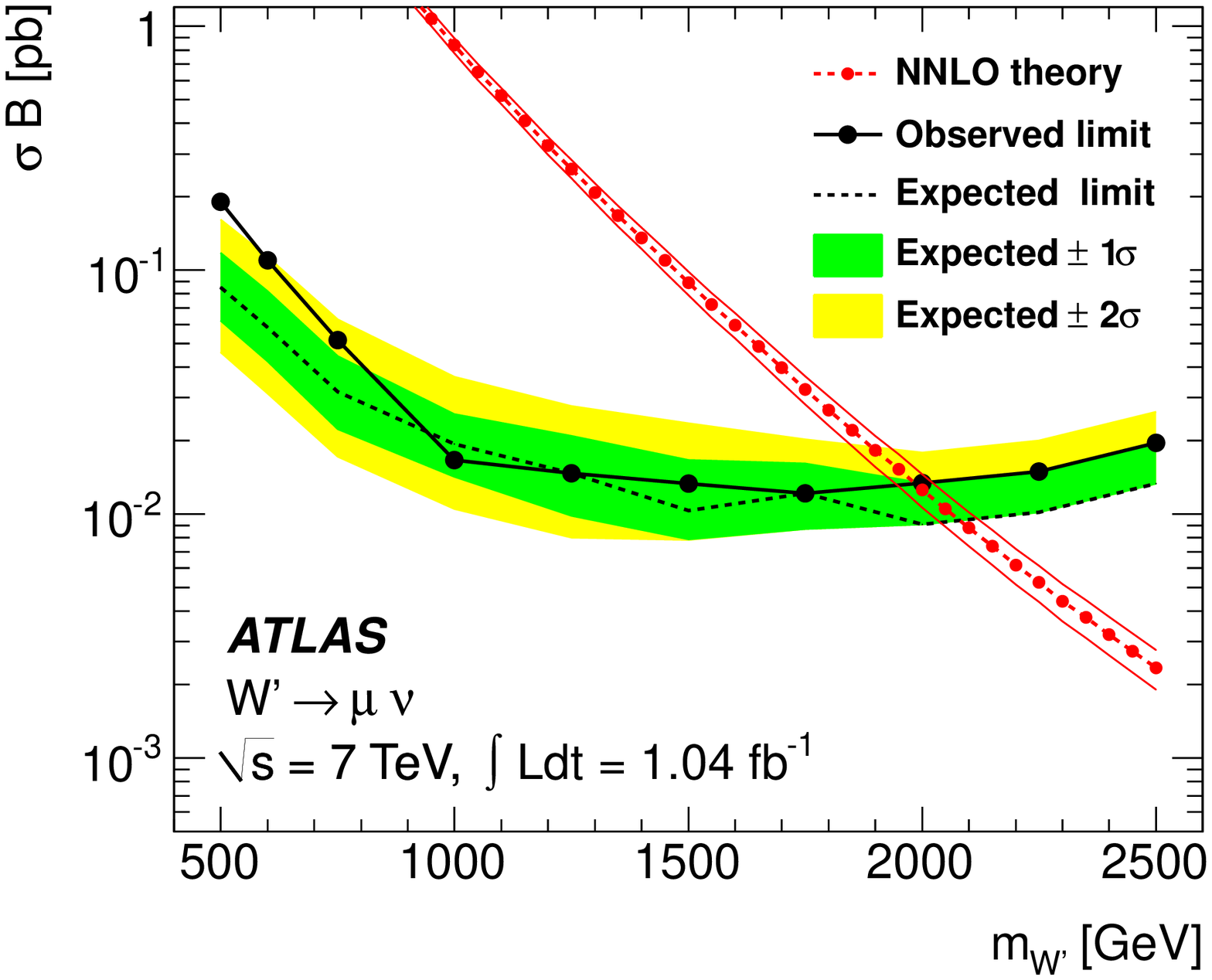}%
\includegraphics[width=80mm]{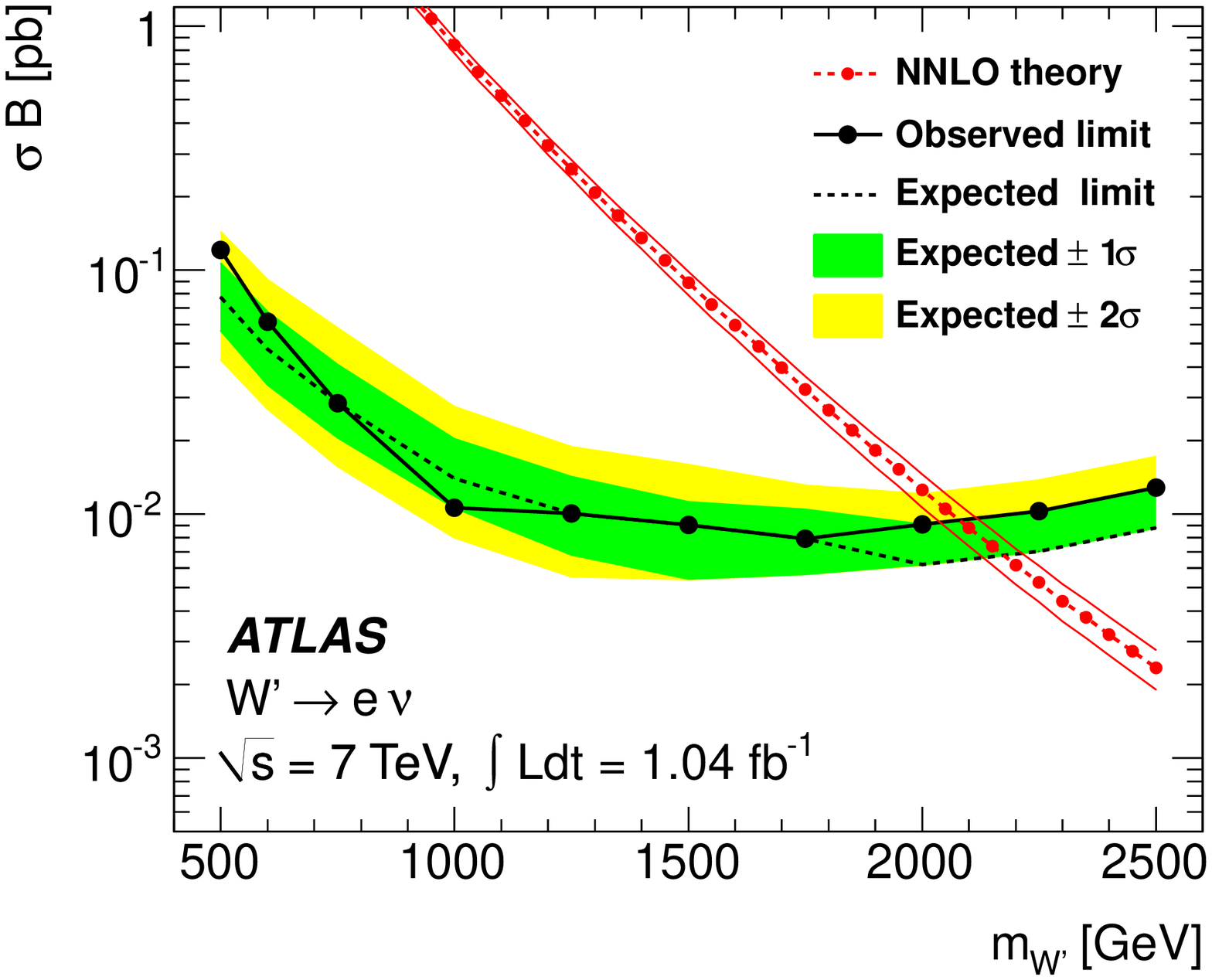}%
\caption{Expected and observed limits on $\sigma B$ for $W^{\prime}\rightarrow \mu\nu$ (left) and $W^{\prime}\rightarrow e\nu$ (right).  The NNLO cross section for a SSM $W^{\prime}$ boson and its uncertainty are also shown.\label{fig:indiv-limits}}
\end{figure}

\begin{figure}
\includegraphics[width=80mm]{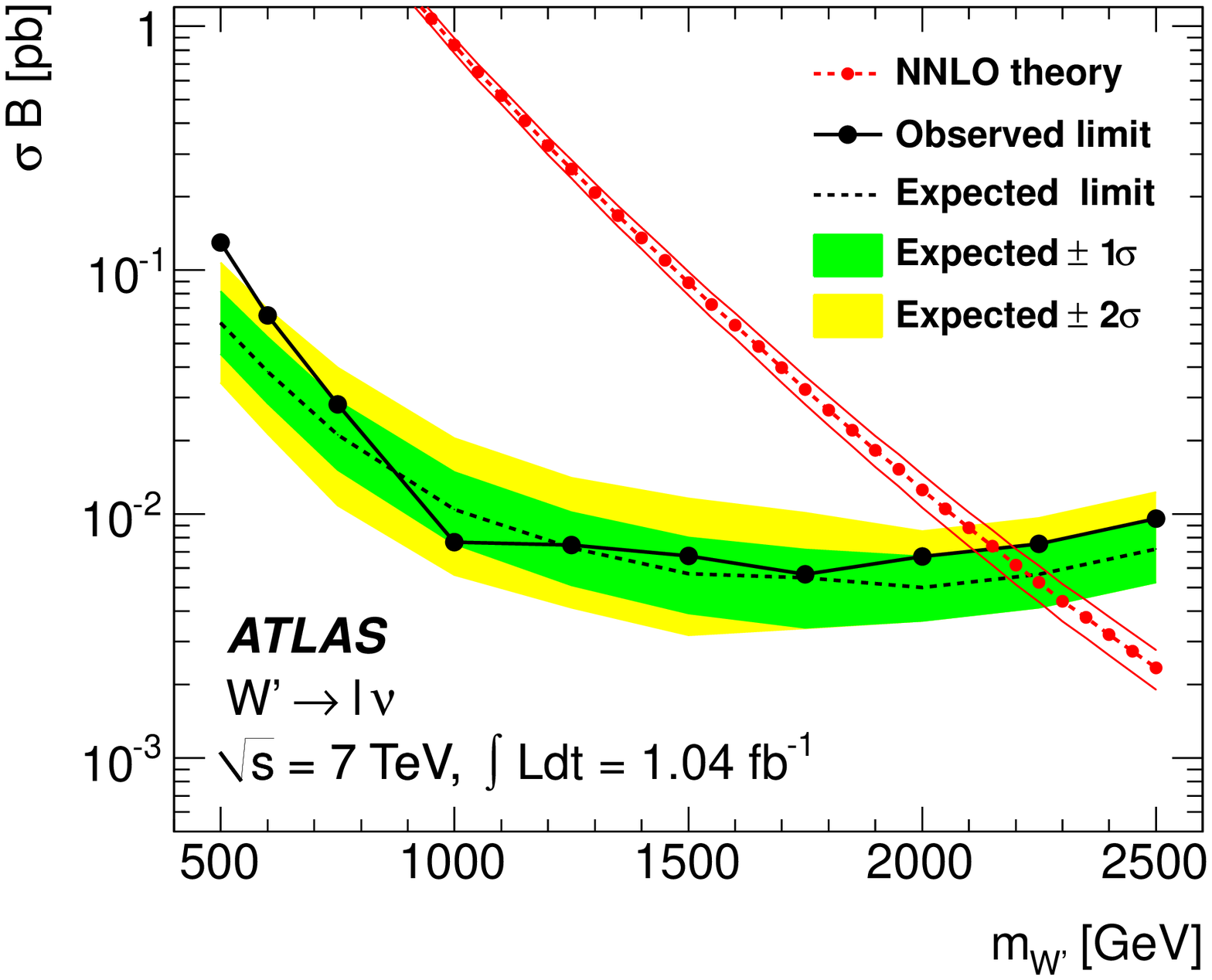}%
\includegraphics[width=80mm]{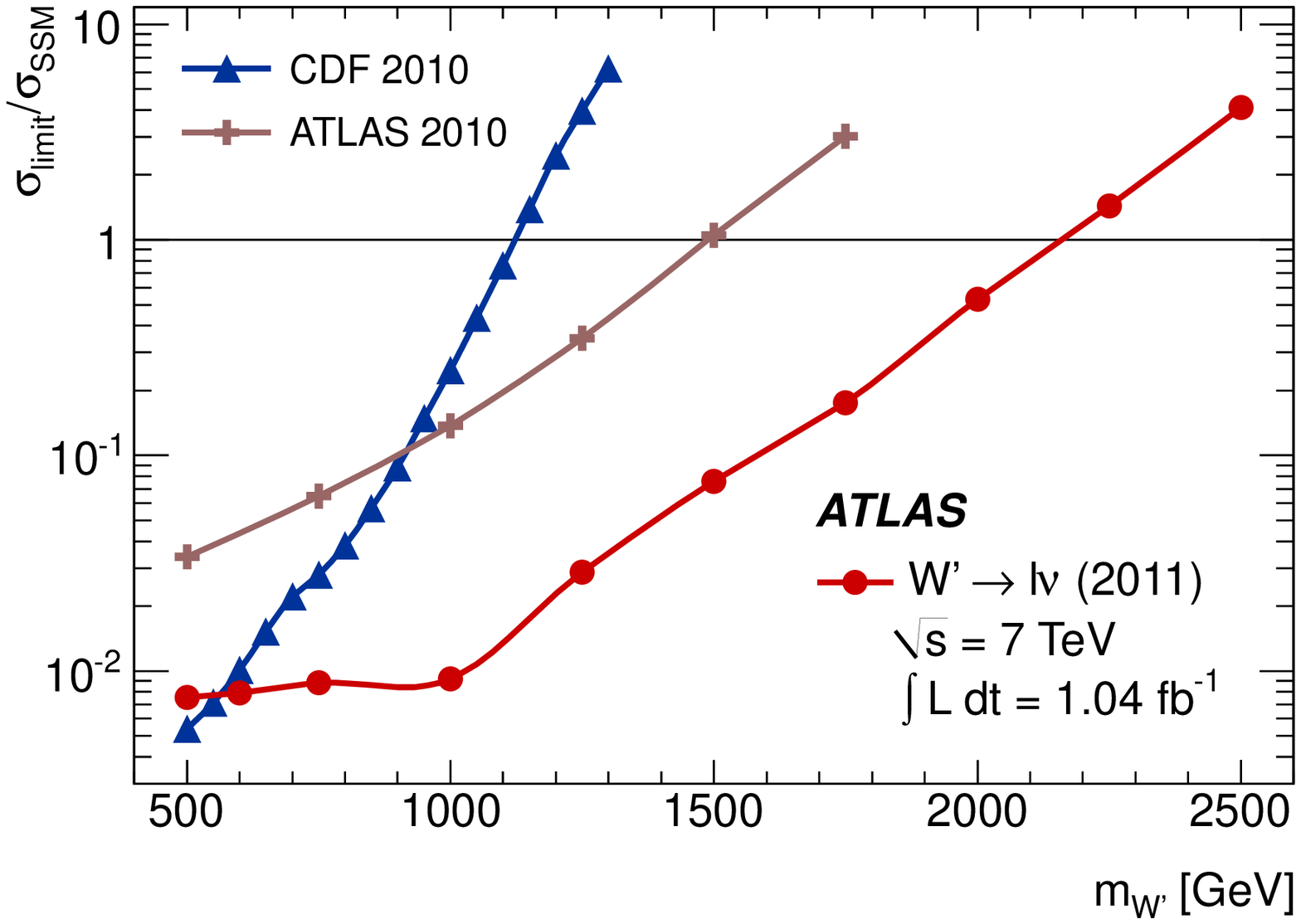}%
\caption{On the left are the expected and observed limits on $\sigma B$ for the combined $W^{\prime}\rightarrow \ell \nu$ ($\ell = e, \mu$ channels) assuming the same branching fraction for both channels.  The NNLO cross section and its uncertainty are also shown.  On the right are the normalized cross section limits ($\sigma_{\textrm{limit}} / \sigma_{\textrm{SSM}}$) for $W^{\prime}\rightarrow \ell \nu$ as a function of mass for this measurement and from CDF~\citep{cdf:Wprime2010}, as well as the previous ATLAS search~\citep{atlas:wprime_2010_pub}.  The cross section calculations assume the $W^{\prime}$ boson has the same couplings as the SM $W$ boson.  The region above each curve is excluded at $95\%$ CL.\label{fig:atlas2011-2-emu}}
\end{figure}

The  limits on the individual channels
are shown in Fig.~\ref{fig:indiv-limits}.
The $W^{\prime}$ boson production is excluded at the $95\%$, shown in 
Fig.~\ref{fig:atlas2011-2-emu}, for masses upto $2.15$~TeV.  Table~\ref{tab:limits_mass} summarizes the lower 
limits on $\sigma B$.  
\begin{table}[ht]
\caption{Lower limits on the SSM $W^{\prime}$ boson mass.
         Rows correspond to the different decay channels ($e\nu$, $\mu\nu$ or both combined) and
         the columns give the expected (Exp.) and observed (Obs.) mass limits.
\label{tab:limits_mass}
}
\begin{center}
\begin{tabular}{c|rr}
\hline
\hline
        &  \multicolumn{2}{c}{$m_{W^{\prime}}$ [TeV]} \\
        &  Exp. & Obs. \\
\hline
 $e\nu$    & 2.17 & 2.08 \\
 $\mu\nu$  & 2.08 & 1.98 \\
 both      & 2.23 & 2.15 \\
\hline
\hline
\end{tabular}
\end{center}
\end{table}
The ratio on the lower limit $\sigma$ for $W^{\prime}$ boson production to the 
SSM production cross section versus $M_{W^{\prime}}$ is shown in Fig.~\ref{fig:atlas2011-2-emu}.  

%
\section{Conclusion \label{sec:conclusion}}
Using $1.04$~fb$^{-1}$ of proton-proton collisions collected at the LHC, ATLAS has searched for a massive boson decaying to a lepton plus missing transverse energy with $m_{\textrm{T}} > 500$~GeV.  No significant excess has been observed, so lower $95\%$ CL limits have been evaluated on the production cross section of a SSM $W^{\prime}$ boson.  This analysis has been submitted to  Physics Review Letters B~\citep{atlas:wp-2011}.
\bigskip 
\begin{acknowledgments}

We thank CERN for the very successful operation of the LHC, as well as the
support staff from our institutions without whom ATLAS could not be
operated efficiently.

We acknowledge the support of ANPCyT, Argentina; YerPhI, Armenia; ARC,
Australia; BMWF, Austria; ANAS, Azerbaijan; SSTC, Belarus; CNPq and FAPESP,
Brazil; NSERC, NRC and CFI, Canada; CERN; CONICYT, Chile; CAS, MOST and
NSFC, China; COLCIENCIAS, Colombia; MSMT CR, MPO CR and VSC CR, Czech
Republic; DNRF, DNSRC and Lundbeck Foundation, Denmark; ARTEMIS, European
Union; IN2P3-CNRS, CEA-DSM/IRFU, France; GNAS, Georgia; BMBF, DFG, HGF, MPG
and AvH Foundation, Germany; GSRT, Greece; ISF, MINERVA, GIF, DIP and
Benoziyo Center, Israel; INFN, Italy; MEXT and JSPS, Japan; CNRST, Morocco;
FOM and NWO, Netherlands; RCN, Norway; MNiSW, Poland; GRICES and FCT,
Portugal; MERYS (MECTS), Romania; MES of Russia and ROSATOM, Russian
Federation; JINR; MSTD, Serbia; MSSR, Slovakia; ARRS and MVZT, Slovenia;
DST/NRF, South Africa; MICINN, Spain; SRC and Wallenberg Foundation,
Sweden; SER, SNSF and Cantons of Bern and Geneva, Switzerland; NSC, Taiwan;
TAEK, Turkey; STFC, the Royal Society and Leverhulme Trust, United Kingdom;
DOE and NSF, United States of America.

The crucial computing support from all WLCG partners is acknowledged
gratefully, in particular from CERN and the ATLAS Tier-1 facilities at
TRIUMF (Canada), NDGF (Denmark, Norway, Sweden), CC-IN2P3 (France),
KIT/GridKA (Germany), INFN-CNAF (Italy), NL-T1 (Netherlands), PIC (Spain),
ASGC (Taiwan), RAL (UK) and BNL (USA) and in the Tier-2 facilities
worldwide.

\end{acknowledgments}

\begin{thebibliography}{9}   
%
\bibitem{atlas:wp-2011}  ATLAS Collaboration, CERN-PH-EP-2011-121 (2011a).
\bibitem{pdg:2010} K. Nakamura et al. (Particle Data Group), J. Phys. G {\bf 37}, 075021 (2010).
\bibitem{ssm} G. Altarelli, B. Mele, and M. Ruiz-Altaba, Z. Phys. C {\bf 45}, 109 (1989).
\bibitem{atlas:wprime_2010_pub} ATLAS Collaboration, Phys. Lett. B {\bf 701}, 50 (2011b).
\bibitem{Moch:2008qy} S. Moch and P. Uwer, Phys. Rev. D {\bf 78}, 034003 (2008), 0804.1476.
\bibitem{Langenfeld:2009tc} U. Langenfeld, S. Moch, and P. Uwer, arXiv:hep-ph/0907.2527 (2009), 0907.2527.
\bibitem{Aliev:2010zk} M. Aliev et al., Comput. Phys. Commun. {\bf 182}, 1034 (2010), 1007.1327.
\bibitem{atlas:photon2010} ATLAS Collaboration, Phys. Rev. D {\bf 83}, 052005 (2011c).
\bibitem{cdf:Wprime2010} T. Aaltonen et al., CDF Collaboration, Phys. Rev. D {\bf 83}, 031102 (2011).
%
%
\end{thebibliography}

\end{document}